\documentclass[twocolumn,showpacs,preprintnumbers]{revtex4}
\usepackage{graphicx}
\usepackage{dcolumn}
\usepackage{bm}
 
\begin{document}
 
\preprint{}
\title{Pairing Instability in a Nematic Fermi Liquid}
\author{Yong Baek Kim and Hae-Young Kee}
\affiliation{Department of Physics, University of Toronto, Toronto, 
Ontario M5S 1A7, Canada}
\date{\today}

\begin{abstract}
Nematic Fermi liquid arises when the system of interacting fermions spontaneously
breaks the rotational symmetry while the translational symmetry is preserved.
We consider a Nematic Fermi liquid of fermions with two distinct quantum numbers.
Two quantum numbers represent the spin degrees of freedom or the layer degrees of 
freedom of the bilayer system. There are two director modes in this system: the in-phase
and out-of-phase modes. We find that the out-of-phase (in-phase) mode mediates 
an attractive (repulsive) interaction between fermions with different quantum numbers.
Pairing instability occurs under certain conditions when the in-phase mode is 
gapped due to a lattice potential. Implications to the stripe phase of Cuprates
and the quantum Hall effect of two dimensional electron systems are discussed.

\end{abstract}
 
\maketitle

\def\be{\begin{equation}}
\def\ee{\end{equation}}
\def\bea{\begin{eqnarray}}
\def\eea{\end{eqnarray}}

\paragraph{Introduction}
The occurrence of the novel anisotropic metallic states
in strongly correlated electron systems has been demonstrated by 
the discoveries of
the stripe phases in Cuprates \cite{tranquada,kivelson99} and the 
anisotropic metallic states in high
Landau levels of two dimensional electron systems in the quantum 
Hall regime \cite{eisenstein99,stormer99,cdw}.
The anisotropic states in these systems are close to the unidirectional 
charge-density-wave states and arise due to 
strong electron-electron interactions. 
In analogy to classical liquid crystals, nematic Fermi liquid state has been
proposed as a primary candidate for the explanation of the quantum liquid 
states mentioned above \cite{fradkin99,macdonald00,wexler01,leo01}. 
In nematic Fermi liquids, the rotational symmetry is 
spontaneously broken while the translational symmetry is still preserved. 
There are two approaches to get nematic Fermi liquid states. One may consider 
first a smectic Fermi liquid state where both the rotational and translational 
symmetries are spontaneously broken \cite{fradkin99,macdonald00}. 
The melting of the smectic state 
via the proliferation of appropriate topological defects will lead to a nematic 
Fermi liquid state. Another approach is to introduce an interaction in ordinary 
Fermi liquid state such that the rotational symmetry will be spontaneously broken 
upon the tunning of the interaction. 

Due to the broken translational symmetry, the smectic state is often visualized
as a liquid version of a uni-directional crystalline 
state \cite{fradkin99,macdonald00,wexler01,leo01}. The low energy 
excitations come from the one-dimensional electronic degrees of freedom that
reside along the stripes in Cuprates or along the edges of two different
quantum Hall states in high Landau levels of two dimensional electron systems.
The existence of the fermionic degrees of freedom adds a major difficulty in
describing the melting of the smectic state to a nematic Fermi liquid state.
So far, no well-defined procedure has been discovered in this direction.
On the other hand, the second approach has been proved to be useful. 
Oganesyan, Kivelson and Fradkin \cite{vadim01} introduced a two-body interaction between 
quadrupolar densities of spinless fermions. Upon the tunning of the interaction, 
the rotational symmetry is spontaneously broken. The Goldstone mode of the nematic 
state is the director mode that corresponds to the angular rotation of the principal
axis of the distorted Fermi surface. Its dispersion relation is highly anisotropic in 
space in contrast to the case of the broken spin symmetries. 
The interaction between the fermions and the anisotropic Goldstone mode leads to 
anisotropic self-energy of the quasiparticles that show non-Landau-Fermi liquid
properties in certain directions. That is, the life time of the quasiparticles
strongly depend on the spatial directions.

In this paper, we follow the second approach and investigate a pairing instability 
of a nematic Fermi liquid of two species of fermions. Two species represent either 
the spin degrees of freedom or the layer degrees of freedom of the bilayer system. 
Introducing an additional interaction between quadrupolar densities of two degrees
of freedom, we find a nematic Fermi liquid state where there are two director modes 
that corresponds to the in-phase and out-of-phase modes of the angular rotations 
of two Fermi surfaces. In general, the out-of-phase mode is gapped while
the in-phase mode remains gapless. 
It is found that the Goldstone mode of the nematic Fermi liquid state of the spinless
fermions always mediates a repulsive interaction. On the other hand, in the case
of the two species of fermions, the out-of-phase mode mediates an attractive
interaction between fermions that belong to different species while the in-phase
mode leads to a repulsive interaction. Since the out-of-phase mode is gapped, one 
expects that the in-phase mode will dominate in the low energy limit and the 
interaction between the fermions will be repulsive. In the presence of an external
potential like a lattice potential, however, the in-phase mode will be also gapped. 
We find in this case that the combined total interaction due to the in-phase and 
out-of-phase modes can be attractive so that there will be a pairing instability 
between the fermions of different species.

\paragraph{Choice of the Hamiltonian}
We consider the following Hamiltonian for a two dimensional fermion
system with two quantum numbers, $\uparrow$ and $\downarrow$.

\begin{eqnarray}
H &=& \int d^2 r \sum_{\alpha}
{\hat \Psi}^{\dagger}_{\alpha} ({\bf r}) \epsilon(-i\nabla) {\hat \Psi}_{\alpha}({\bf r}) \cr
&+& \int d^2 r d^2 r' F_2 ({\bf r}-{\bf r'})
\sum_{\alpha} {\rm Tr}[{\hat Q}_{\alpha} ({\bf r}) {\hat Q}_{\alpha} ({\bf r'})] \cr
&+& \int d^2 r d^2 r' G_2 ({\bf r}-{\bf r'})
\sum_{\alpha \not= \beta} {\rm Tr}[{\hat Q}_{\alpha} ({\bf r}) {\hat Q}_{\beta} ({\bf r'})] \ ,  
\end{eqnarray}
where $\alpha, \beta = \uparrow, \downarrow$, ${\hat \Psi}_{\alpha}$ is 
an annihilation operator of the fermion with ``spin'' $\alpha$,
and $\epsilon (-i\nabla)$ is the kinetic energy operator.
Here the matrix quadrupolar density of the fermions is defined as
\begin{equation}
{\hat Q}_{\alpha} ({\bf r}) = - {1 \over k_F^2} 
{\hat \Psi}^{\dagger}_{\alpha} ({\bf r}) 
\left (\matrix{\partial^2_x - \partial^2_y & 2 \partial_x \partial_y \cr
2 \partial_x \partial_y & \partial^2_y - \partial^2_x \cr} \right )
{\hat \Psi}_{\alpha} ({\bf r})
\end{equation}
The interactions, $F_2 ({\bf q})$ between the fermions with the same ``spin''
and $G_2 ({\bf q})$ between the fermions with different ``spin'', are given by
\begin{equation}
F_2 ({\bf q}) = \frac{1}{F^{-1}_2+\kappa_1 q^2}, \ \ \
G_2 ({\bf q}) = \frac{1}{G^{-1}_2+\kappa_2 q^2} \ .
\end{equation}
The details of the form of the interaction do not matter as far as the short 
range interaction is concerned. 

Using a Hubbard-Stratonovich transformation, we 
can write the effective Lagrangian as
\begin{eqnarray}
{\cal L} &=& \int d^2 r \sum_{\alpha}
\Psi^{\dagger}_{\alpha} ({\bf r},\tau) (\partial_{\tau} + \epsilon(-i\nabla) - \mu) 
\Psi_{\alpha}({\bf r},\tau) \cr
&+& \int \frac{d^2 q}{(2\pi)^2} F_2 ({\bf q}) [1 + g({\bf q})] \ 
{\rm Tr}[Q_{+} ({\bf q},\tau) N_{+} ({\bf -q},\tau)] \cr 
&+& \int \frac{d^2 q}{(2\pi)^2} F_2 ({\bf q}) [1 - g({\bf q})] \
{\rm Tr}[Q_{-} ({\bf q},\tau) N_{-} ({\bf -q},\tau)] \cr 
&+& \int \frac{d^2 q}{(2\pi)^2} F_2 ({\bf q}) [1 + g({\bf q})] 
{\rm Tr}[N_{+} ({\bf q},\tau) N_{+} ({\bf -q},\tau)] \cr
&+& \int \frac{d^2 q}{(2\pi)^2} F_2 ({\bf q})  [1 - g({\bf q})] 
{\rm Tr}[N_{-} ({\bf q},\tau) N_{-} ({\bf -q},\tau)], 
\end{eqnarray}
where $Q_{\pm}({\bf q},\tau) = Q_{\uparrow}({\bf q},\tau) 
\pm Q_{\uparrow}({\bf q},\tau)$ with
\begin{eqnarray}
Q_{\alpha} ({\bf q},\tau) = 
\sum_{\bf k} \Psi^{\dagger}_{\alpha, {\bf k}-{\bf q}}(\tau)
\left (\matrix{\cos (2 \theta_{\bf k}) & \sin (2 \theta_{\bf k}) \cr 
\sin (2 \theta_{\bf k}) & -\cos (2 \theta_{\bf k}) \cr} \right )
\Psi_{\alpha, {\bf k}}(\tau)
\end{eqnarray}
and $\theta_{\bf k}$ is the angle between ${\bf k}$ and the $x$-axis.
Here $g({\bf q}) = G_2({\bf q})/F_2({\bf q})$ and $N_{\pm}({\bf q},\tau)$ 
are the Hubbard-Stratonovich matrix fields given by
\begin{equation}
N_{\pm}({\bf q},\tau) = \left (\matrix{ N_{\pm, 1}({\bf q},\tau) 
& N_{\pm, 2}({\bf q},\tau) \cr
N_{\pm, 2}({\bf q},\tau) 
& -N_{\pm, 1}({\bf q},\tau) \cr}\right ).
\end{equation}  
Notice that, at the saddle point of the action, 
$N^{sp}_{\pm} ({\bf q})$ corresponds to
$\langle {\hat Q}_{\pm}({\bf q})\rangle$. 
 
\paragraph{Saddle Point and Gaussian Fluctuations}
Now one can integrate out the fermionic degrees of freedom and
get an effective action in terms of $N_{+}({\bf q})$ and $N_{-}({\bf q})$.
In the putative nematic liquid state, there should be a uniform saddle point 
solution $N^{sp}_{\pm} \not= 0$. Notice that non-zero
$N^{sp}_{+, 1}$ ($N^{sp}_{-, 1}$) 
will lead to the symmetric (antisymmetric) distortions of
the $\uparrow$ and $\downarrow$ Fermi surfaces.
The finite value of $N^{sp}_{+, 2}$  
($N^{sp}_{-, 2}$) determines the amount of the 
symmetric (antisymmetric) rotations of the principal axis of 
the $\uparrow$ and $\downarrow$ Fermi surfaces. 
The saddle point solution depends on the coefficient of the
quadratic term in the long wavelength (${\bf q} \rightarrow 0$) expansion
of the Landau-Ginzburg free energy.
\begin{eqnarray}
{\cal F} (N_+,N_-) = M_+ {\rm Tr}[N^2_+] + M_- {\rm Tr}[N^2_-] + \cdots,
\end{eqnarray}
where $\cdots$ represents the quartic and higher order terms.
One can show that
\begin{eqnarray}
M_+ &=& 2(1+g) F_2 + (1+g)^2 N(0) F^2_2 \cr
M_- &=& 2(1-g) F_2 + (1-g)^2 N(0) F^2_2, 
\end{eqnarray}
where $g = g({\bf q}=0)$ and $N(0)$ is the density of states
at the Fermi level. We will choose $F_2 > 0$ and $G_2 < 0$ 
such that $M_+ < 0$. One can see that $M_- > 0$ in this case.
Then the saddle point solution, $N^{sp}_+ \not= 0$ and $N^{sp}_- = 0$,
represents the symmetric distortions and rotations of the $\uparrow$
and $\downarrow$ Fermi surfaces. 

The Gaussian fluctuations of 
$N_{\pm}({\bf q}) = N^{sp}_{\pm} + \delta N_{\pm}({\bf q})$
about the nematic liquid ground state and their interaction with the 
fermions can be written as ($a = 1,2$)
\begin{eqnarray}
{\cal L}_{\rm eff} &=& {\cal L}_{\rm sp} + {\cal L}_{\rm int} + {\cal L}_{g}, \cr
{\cal L}_{\rm int} &=& \int \frac{d^2 q}{(2\pi)^2} \{ F_2 (1+g) 
{\rm Tr}[Q_{+} ({\bf q},\tau) \delta N_{+} ({\bf -q},\tau)] \cr
&&\hskip 0.5cm + F_2 (1-g)
{\rm Tr}[Q_{-} ({\bf q},\tau) \delta N_{-} ({\bf -q},\tau)] \} + c.c. \cr
{\cal L}_{g} 
&=& \int \frac{d^2 q}{(2\pi)^2} \
\delta N_{+,a} ({\bf q},\tau) D^{-1}_{+,ab} ({\bf q},\tau) 
\delta N_{+,b} ({\bf -q},\tau) \cr
&+& \int \frac{d^2 q}{(2\pi)^2} \ 
\delta N_{-,a} ({\bf q},\tau) D^{-1}_{-,ab} ({\bf q},\tau) 
\delta N_{-,b} ({\bf -q},\tau),
\label{effaction}  
\end{eqnarray}
where ${\cal L}_{sp}$ describes the mean field nematic liquid ground state.

The propagators of the Gaussian fluctuations, $D^{-1}_{\pm,ab}$ can
be obtained at the Gaussian level (Random Phase Approximation) as follows.
\begin{equation}
D^{-1}_{\pm,ab} ({\bf q},i\nu_n) = 
\Delta_{\pm,ab} + (1 \pm g)^2 F^2_2 \Pi_{ab} ({\bf q},i\nu_n),
\end{equation}
with
\begin{equation}
\Delta_{\pm,ab} = \left (\matrix{\Delta_{\pm,1} & 0 \cr 
0 & \Delta_{\pm,2} \cr} \right ),
\end{equation}
and
\begin{eqnarray}
\Pi_{ab} ({\bf q},i\nu_n) &=& N(0) 
\int \frac{d\theta}{2\pi} \frac{s}{s-\cos(\theta-\phi)} \cr
&\times& \left (\matrix{2\cos^2(2 \theta) & \sin(4\theta) \cr 
\sin (4\theta) & 2\cos^2(2 \theta) \cr} \right ),
\end{eqnarray}
where $\phi$ is the angle between ${\bf q}$ and the $x$-axis.
Here $\nu_n = 2n \pi T$ is the Matsubara frequency and $s = i\nu_n/v_Fq$.
$\Delta_{\pm,1}$ and $\Delta_{\pm,2}$ represent the gap in each 
collective mode, that depend on the interaction strength, but
their precise forms are not important in the following analysis.
In the nematic liquid state, for generic interactions without an external
perturbation, we expect that there should be one gapless Goldstone mode, 
the in-phase director mode, associated with $\delta N_{+,2}$.
That is, $\Delta_{+,2} = 0$. 
If the system experiences a weak square lattice potential or a weak
quadrupolar external perturbation, even the in-phase director mode will
be gapped.
As will be shown later, both the in-phase ($\delta N_{+,2}$) and 
out-of-phase ($\delta N_{-,2}$) director modes will be important 
in the analysis of the pairing instability. The distortion modes,
$\delta N_{+,1}$ and $\delta N_{-,1}$, are gapped even in the absence
of the interaction between $\uparrow$ and $\downarrow$ fermions, thus
one expects that the gap of these modes will be even bigger.
We will drop the distortion modes in the analysis from now on.

The propagators of two director modes in the long wavelength and low
energy limits can be obtained as
\begin{equation}
D^{-1}_{\pm,22} = 
\Delta_{\pm,2} + \kappa_{\pm} q^2 + 
2 (1\pm g)^2 N(0) F^2_2 \frac{|\nu_n|}{v_F q} \sin^2 (2\phi)     
\end{equation}   
where, in the absence of the external potential, $\Delta_{+,2} = 0$ 
and $\Delta_{-,2} > 0$. When there is a weak quadrupolar external potential, 
we expect $\Delta_{-,2} > \Delta_{+,2} \not= 0$. Here $\kappa_+$ and $\kappa_-$
are the stiffness coefficients and their detailed dependence on the
interaction strength will not be important.

\begin{widetext}
\paragraph{Effective Two-Body Interaction}
The fermions with $\uparrow$ and $\downarrow$ quantum numbers interact
each other by exchanging two director modes just like the case of the
phonon mediated electron-electron interaction. The effective two-body
interaction between fermions can be obtained by integrating out the 
director modes in the effective action given by Eq.\ref{effaction}
and using the following form of the coupling between the director modes
and the fermions,
\begin{equation}
(1+g) \delta N_{+,2} ({\bf q})
\sum_{\bf k}  
\sin(2\theta_{\bf k}) \Psi^{\dagger}_{\alpha,{\bf k}-{\bf q}} 
\Psi_{\alpha,{\bf k}} 
+ (1-g) \delta N_{-,2} ({\bf q})
\sum_{\bf k} 
\sin(2\theta_{\bf k}) \Psi^{\dagger}_{\alpha,{\bf k}-{\bf q}}
\sigma^z_{\alpha \beta} \Psi_{\beta,{\bf k}}.
\end{equation} 
After integrating out the director modes, the effective interaction
between fermions is obtained as
\begin{eqnarray}
S'_{\rm eff} &=& \sum_{{\bf q},\nu_n}
\sum_{{\bf k},\omega_m} \sum_{{\bf k'},\omega'_{m}} \sum_{\alpha}
V_1({\bf k},{\bf k'},{\bf q},i\nu_n)
\Psi^{\dagger}_{\alpha,{\bf k}-{\bf q}}(\omega_m-\nu_n)  
\Psi^{\dagger}_{\alpha,{\bf k'}+{\bf q}}(\omega'_m+\nu_n) 
\Psi_{\alpha,{\bf k'}}(\omega'_m) \Psi_{\alpha,{\bf k}}(\omega_m) \cr
&+& \sum_{{\bf q},\nu_n}
\sum_{{\bf k},\omega_m} \sum_{{\bf k'},\omega'_{m}} 
\sum_{\alpha \not= \beta}
V_2({\bf k},{\bf k'},{\bf q},i\nu_n)
\Psi^{\dagger}_{\alpha,{\bf k}-{\bf q}}(\omega_m-\nu_n) 
\Psi^{\dagger}_{\beta,{\bf k'}+{\bf q}}(\omega'_m+\nu_n) 
\Psi_{\beta,{\bf k'}}(\omega'_m) \Psi_{\alpha,{\bf k}}(\omega_m),
\end{eqnarray}  
where
\begin{eqnarray}
V_1({\bf k},{\bf k'},{\bf q},i\nu_n) &=& F^2_2 [(1+g)^2 D_{+,22}({\bf q},i\nu_n) 
+ (1-g)^2 D_{-,22}({\bf q},i\nu_n)] \sin(2\theta_{\bf k}) \sin(2\theta_{\bf k'}) \cr
V_2({\bf k},{\bf k'},{\bf q},i\nu_n) &=& F^2_2 [(1+g)^2 D_{+,22}({\bf q},i\nu_n) 
- (1-g)^2 D_{-,22}({\bf q},i\nu_n)] \sin(2\theta_{\bf k}) \sin(2\theta_{\bf k'}). 
\end{eqnarray}
\end{widetext}

\paragraph{Cooper Channel and Pairing Instability} 
Notice that $V_1$ and $V_2$ correspond to the interactions between the fermions 
with the same spin and with different spins respectively.
In the Cooper channel, ${\bf k'}=-{\bf k}$ and $\theta_{\bf k'}=\theta_{\bf k}+\pi$,
$V_1$ is always repulsive while $V_2$
can be attractive if $|g| = |G_2|/F_2 > 1$. In the long wavelength and low energy
limit, $V_2$ in the Cooper channel can be written as (here 
$\Delta_{+} \equiv \Delta_{+,2}$ and $\Delta_{-} \equiv \Delta_{-,2}$)
\begin{equation}
V_2({\bf k},-{\bf k},{\bf q},i\nu_n) = - 
\frac{A \sin^2(2\theta_{\bf k})}{C + B \frac{|\nu_n|}{v_F q} \sin^2 (2\phi)}, 
\end{equation}  
where
\begin{eqnarray}
A &=& F^2_2 \ [ \ 2|g|(\Delta_+ + \Delta_-) - (|g|^2+1)(\Delta_- - \Delta_+) \ ] \cr
B &=& 2 N(0) F^2_2 \ [(|g|^2+1)(\Delta_+ + \Delta_-) - 2|g|(\Delta_- - \Delta_+)] \cr
C &=& \Delta_+ \Delta_-. 
\end{eqnarray}  
One can show that $V_2 < 0$ if the following condition is satisfied.
\begin{equation}
\frac{(\sqrt{\Delta_-} - \sqrt{\Delta_+})^2}{\Delta_- - \Delta_+} 
< |g| <
\frac{(\sqrt{\Delta_-} + \sqrt{\Delta_+})^2}{\Delta_- - \Delta_+}  
\end{equation}

\paragraph{Gap Equation and Nematic Superconductor}
Using the attractive interaction $V_2$, the gap equation for 
$\Delta ({\bf k}) = \langle \Psi_{\uparrow, {\bf k}}
\Psi_{\downarrow, -{\bf k}} \rangle$ can be written as
\begin{equation}
\Delta ({\bf k},i\nu_n) = - T \sum_{{\bf k'},\nu'_n} 
\frac{V_2 ({\bf k},-{\bf k},{\bf k}-{\bf k'},i\nu_n-i\nu'_n) 
\Delta({\bf k'},i\nu'_n)}
{(\nu'_n)^2 + \xi^2_{\bf k'}+ \Delta^2({\bf k'},i\nu'_n)}
\end{equation}
Assuming that $\Delta ({\bf k},i\nu_n)$ depends mostly on the angle
$\theta_{\bf k}$, we can perform the integral over 
$\xi_{\bf k'}=\epsilon_{\bf k'}-\mu$ and get the following results at T=0.
\begin{widetext}
\begin{equation}
\Delta (\theta_{\bf k},i\nu) = N(0) \int \frac{d\theta}{2\pi} 
\int \frac{d\nu'}{2\pi} \frac{\Delta(\theta_{\bf k}+\theta,i\nu')}
{\sqrt{(\nu')^2 + \Delta^2(\theta_{\bf k}+\theta,i\nu')}}
\frac{A \sin^2(2\theta_{\bf k})}{C + B \frac{|\nu-\nu'|}{v_F k_F |\theta|}
\sin^2(2\theta_{\bf k})},
\label{gap}
\end{equation}
\end{widetext}
where $\theta = \theta_{\bf k} - \theta_{\bf k'}$. We also use 
$q \approx k_F |\theta|$ and $\sin^2(2\phi) \approx 
\sin^2(2\theta_{\bf k})$ in the long wavelength and low frequency limit.
One can easily see that $\Delta (\theta_{\bf k},i\nu_n)$ vanishes
at $\theta_{\bf k} = 0, \pm {\pi \over 2}, \pi$. 
It can be shown that the frequency dependence in $\Delta (\theta_{\bf k},i\nu_n)$
is weak near $\theta_{\bf k} = 0, \pm {\pi \over 2}, \pi$.
When $\Delta(\theta_{\bf k}) \sin^2(2\theta_{\bf k}) \ll E_{\rm eff}$, we get 
\begin{equation}
\Delta(\theta_{\bf k}) \approx
\frac{E_{\rm eff}}{\sin^2(2\theta_{\bf k})} 
\exp \left ( 
-\frac{1}{N(0) V_{\rm eff} \sin^2(2\theta_{\bf k})} \right ),
\end{equation}
where 
\begin{eqnarray}
E_{\rm eff} &=& \frac{2 \theta_c}{e} \frac{C}{B} v_F k_F \cr
V_{\rm eff} &=& \frac{\theta_c}{\pi} \frac{A}{C}.
\end{eqnarray}
Here $\theta_c$ is the cut-off in the integral over the
angle $\theta$ in Eq.\ref{gap}. 
This is the asymptotically correct solution near
$\theta_{\bf k} = 0, \pm {\pi \over 2}, \pi$.
Thus the gap is exponentially suppressed in the four
directions. We have to mention, however, that the whole angular 
dependence of the gap will depend on the underlying lattice 
structure.

\paragraph{Summary and Conclusion}
We study the pairing instability in nematic Fermi liquid states where
the orientational order is spontaneously broken due to an electron-electron
interaction. In the case of the fermions with no internal quantum number
(e.g. spin polarized electrons), the director mode -- the goldstone mode
of the nematic order -- always mediates a repulsive interaction in the
Cooper channel. On the other hand, when the fermions carry two different
quantum numbers (e.g. spin or the layer index of the bilayer system),
two independent director modes exist. The out-of-phase (in-phase) mode
mediates an attractive (a repulsive) interaction in the Cooper channel.
When an external potential is applied to the system so that the in-phase
mode becomes gapped, the total effective interaction can be attractive
under certain conditions. We showed that this leads to a pairing instability.
The resulting pairing gap has a remarkable structure. There are four nodes
where the gap vanishes and the gap is exponentially suppressed near the nodes.
The angular dependence near the nodes cannot be described by an analytic
expansion in the angle around the nodes.  

Our results may have the following implications to the physics of 
Cuprates and quantum Hall effect.
1) The occurrence of the superconductivity in the stripe phase of
Cuprates has been a subject of great interest. In particular, the question
of whether the stripes (or some dynamic generalization) are in favor of or 
against superconductivity has not been clearly answered \cite{kivelson99}. 
Our work on the pairing instability of a nematic Fermi liquid state will shed some
light on this question especially because the existence
of the nematic order and its (pseudo-)Goldstone mode is crucial 
for the effective attractive interaction and pairing instability.
2) In the lowest Landau level, the Fermi-liquid-like state \cite{HLR}
of the filling factor $\nu=1/2$ can be described as a Fermi liquid state 
of the quasiparticles called composite fermions \cite{jain}. 
There have been
a number of examples of the paired quantum Hall states in the bilayer
generalization, where the ground state can be described as
a paired state of the composite fermions \cite{read00,bonesteel,ybkim}. 
In the half-filled
high Landau levels, the ground state is a highly anisotropic metallic
state very close to a unidirectional charge density wave \cite{cdw}. 
If one can
consider the nematic Fermi liquid state of the composite fermions
as a candidate for the anisotropic state mentioned above \cite{vadim01}, 
one would expect that an unusual paired quantum Hall state may occur in high 
Landau levels in the bilayer system, arising from the pairing 
instability of the underlying nematic Fermi liquid states of 
the composite fermions.  

{\it Acknowledgments}
We thank Steve Kivelson, Leo Radzihovsky, and Vadim Oganesyan 
for useful discussions.
This work was supported in part by Canadian Institute for 
Advanced Research (H.Y.K. and Y.B.K.) and Alfred P. Sloan Foundation (Y.B.K.).

\end{document}